# How long single-photon detectors stay in quantum superpositions during detection according to the Diósi-Penrose criterion

*Garrelt Quandt-Wiese*
*Schlesierstr. 16*
*64297 Darmstadt*
*Germany*
*garrelt@quandt-wiese.de*
*www.quandt-wiese.de*

For special single-photon detectors that are isolated from their environment during detection (so-called indirect detectors), it is investigated how long they stay in a superposition of a photon-detected and a no-photon-detected state according to the Diósi–Penrose criterion for wavefunction collapse. To suppress interactions with the environment during detection, the avalanche photodiodes of the indirect detectors are biased using plate capacitors rather than conventional voltage sources, and the detection outcome is read out a sufficient time after the superposition in the detector has reduced. For the analysis, the Diósi-Penrose criterion is applied to solids in quantum superpositions that are slightly displaced relative to each other or have slightly different expansions in the superposed states, where both the parameter-free Diósi-Penrose model and Diósi's version, in which the microscopic mass distribution is spatially averaged, are discussed. It is shown that indirect single-photon detectors can be constructed in such a way that they remain in superposition for seconds. It is proposed to use indirect detectors for the generation of mirror superpositions with the help of piezo-actuators, where the superposed mirror can have a displacement of about 50 Å for approximately half a microsecond. Even though the superposed mirror states generated in this way are decoherent superpositions (improper mixtures) and therefore cannot be detected by conventional methods, their generation opens new perspectives for probing wavefunction collapse.

.

## 1. Introduction

In the conventional interpretation of quantum theory, measurements on superposed states, such as single-photon detections, lead to an immediate collapse of the superposition. This view must be refined if one assumes that collapse is not caused by the act of measurement but is governed by objective laws. In this case, the internal states of the detector that lead to the final measurement outcome must be considered, as they become part of the initial quantum superposition. This raises the question of how long these internal states of the detector stay in superposition before the detector reduces to the final measurement result.
In this work, we analyse this question for single-photon detectors based on avalanche photodiodes. The discussion shows that not only the photodiode but also all components of the detector, such as its power supply, and even a human observer viewing the detector's result must be considered when taking part in the detection process. This shows that our question can only be meaningfully addressed if one examines specially designed detectors that do not interact with their environment during measurement and whose measurement outcome is read out afterwards, when the detector already has undergone reduction. These detectors, which we refer to here as *indirect detectors*, are characterized by "*measurement after reduction*" rather than "*reduction by measurement*".

For the discussion of the proposed question, a collapse model is needed. There is a general uncertainty in this context, since to date no experimentally verified collapse model exists, and the proposed models are unable to resolve the conflict between the experimentally established nonlocal nature of wavefunction collapse and the local nature of relativity. For the analysis in this work, the so-called Diósi–Penrose criterion (section 2) is chosen, which has already been used to assess experimental proposals for measuring wavefunction collapse [1-9]. The Diósi–Penrose criterion [10,11] is based on the gravity-related models of Diósi [12,13] and Penrose [14,1]. An alternative is provided by the Dynamical Reduction Models [15], among which the model of Continuous Spontaneous Localizations (CSL) has already been used for the discussion of concrete experiments [16] and would also be of interest for the analysis addressed here. When using the CSL model, an additional difficulty arises in that the fundamental parameters of the model (the collapse rate $\lambda$ and the collapse length $r_c$) are not exactly known and must be assumed. This difficulty does not arise for the Diósi–Penrose criterion, since in this case the collapse rate is determined by gravity, and the collapse length, whose introduction is only necessary in Diósi's model (section 2.1), is expected to lie in a range in which it has hardly any influence on the predictions (see section 3.4).

A major problem in the experimental verification of superposed states for measuring collapse is the decoherence that inevitably occurs between the states. The phenomenon of decoherence cannot be distinguished from collapse, since both effects lead to the disappearance of interference between the states, and decoherence usually sets in much earlier than the collapse itself. Since this will also be the case for the superposed states of the indirect single-photon detector discussed in section 5, the superpositions of detector states discussed here must always be understood as decoherent superpositions, also referred to as improper mixtures, which cannot be detected by means of quantum interference. Even though the question discussed here - how long such decoherent superpositions can persist - may at first appear purely academic, since they cannot be detected using standard methods, we will give in section 6 an outlook that generating and investigating specific decoherent superpositions with indirect detectors might help to advance our understanding of wavefunction collapse.

Since single-photon detectors are made of solids, this work first develops in section 3 an extensive collection of formulas to calculate the lifetimes of superposed solids according to the Diósi–Penrose criterion, in which the superposed states may be slightly displaced relative to one another or may have slightly different spatial extensions. In the analysis, the microscopic mass distribution of the solid, which is determined by the atomic nuclei and whose spatial variations are caused by the excited acoustical phonons in the solid, is taken into account, and differences between Diósi's and Penrose's models are discussed. Even though the indirect detectors considered in section 5 are discussed exclusively for room temperature, the foundations developed here for solids can also be used for the discussion of experiments at very low temperatures. In section 4, the results for solids are applied to electrical components such as capacitors, resistors, wires, and piezo-actuators, which are required for the discussion of the indirect detectors in section 5.

In section 5.1, a design for an indirect single-photon detector is proposed, and in section 5.2 it is shown that this detector can be constructed in such a way that it stays at detection in a superposition of a photon-detected and a no-photon-detected state for up to seconds. Since detectors constructed in this way have, due to their long lifetimes, hardly any influence on the collapse of a superposition, they can be used to deliberately generate decoherent superpositions. An example is given in section 5.3, in which an indirect single-photon detector uses a piezo-actuator to displace a mirror that can be part of an interferometer. In section 5.4, we then show how the lifetimes of indirect single-photon detectors can be deliberately controlled using such a setup. The paper ends with an outlook in section 6 on how the generation of decoherent mirror superpositions in interferometers might be used to measure wavefunction collapse in the future.

## 2. Diósi-Penrose criterion

At the Diósi-Penrose criterion, the mean lifetime of a superposition $T_G$ can be estimated with a characteristic gravitational energy between the superposed states $E_{DP}$ by:

$$T_G \approx \frac{\hbar}{\gamma E_{DP}} . \tag{1}$$

The energy $E_{DP}$, which we denote here as the *Diósi-Penrose energy*, depends on the difference of the mass density distributions $\rho_1(\boldsymbol{x})$ and $\rho_2(\boldsymbol{x})$ of the superposed states and is given by [17-19,1]:

$$E_{DP} = \frac{G}{2} \int d^3\boldsymbol{x} d^3\boldsymbol{y} \frac{(\rho_2(\boldsymbol{x}) - \rho_1(\boldsymbol{x}))(\rho_2(\boldsymbol{y}) - \rho_1(\boldsymbol{y}))}{|\boldsymbol{x} - \boldsymbol{y}|} \quad , \tag{2}$$

where $G$ is the gravitational constant. The factor $\gamma$ in (1) is a dimensionless factor, which is expected to be of the order of one [1]. In all subsequent calculations, $\gamma$ is assumed to be one and therefore not listed explicitly.
The Diósi-Penrose energy can also be written as follows, where $\Phi_i(\boldsymbol{x})$ is the gravitational potential corresponding to the mass distribution $\rho_i(\boldsymbol{x})$ ($\Phi_i(\boldsymbol{x}) = -\int d^3\boldsymbol{y} \frac{\rho_i(y)}{|y-x|}$).)

$$E_{DP} = -\frac{1}{2} \int d^3\mathbf{x} \, (\rho_2(\mathbf{x}) - \rho_1(\mathbf{x}))(\Phi_2(\mathbf{x}) - \Phi_1(\mathbf{x})) \quad . \tag{3}$$

From this illustration it follows that the Diósi-Penrose energy is minus the gravitational self-energy resulting from the difference of the mass density distributions $\rho_2(\boldsymbol{x}) - \rho_1(\boldsymbol{x})$ of the superposed states. For some calculations it is helpful to write the Diósi-Penrose energy in the following form [1]:

$$E_{DP} = \frac{1}{8\pi G}\int d^3\boldsymbol{x}|g_1(\boldsymbol{x}) - g_2(\boldsymbol{x})|^2 \quad , \tag{4}$$

where $g_i(\boldsymbol{x})$ are the gravitational fields resulting from the gravitational potentials $\Phi_i(\boldsymbol{x})$ ($g_i(\boldsymbol{x}) = -\nabla\Phi_i(\boldsymbol{x})$). From equation (4) it follows that the Diósi-Penrose energy is always positive.

### 2.1. Diósi-Penrose model with smeared mass-density operator

An important difference in Diósi's model compared to Penrose's approach is that there is a need to modify the mass-density operator to avoid divergences in the master equation for the evolution of the density matrix. This can be achieved by introducing a characteristic smearing radius $R_{sm}$ for the calculation of the mass density with the mass-density operator $\hat{\rho}(\boldsymbol{x})$ as follows [12,13]:

$$\hat{\rho}(\boldsymbol{x}) = \sum_i m_i\,\delta(\boldsymbol{x} - \hat{\boldsymbol{x}}_i) \quad \Rightarrow \quad \hat{\rho}(\boldsymbol{x}) = \sum_i \frac{m_i}{\frac{4\pi}{3}R_{sm}^3}\Theta(R_{sm} - |\boldsymbol{x} - \hat{\boldsymbol{x}}_i|) \ , \tag{5}$$

where $m_i$ and $\hat{\boldsymbol{x}}_i$ are the mass and the position operator of the i's particle, respectively. The choice of the smearing radius $R_{sm}$ determines how much the total energy of a system increases over time. To avoid unrealistically large increases, it was suggested to choose $R_{sm}$ in the order of $10^{-5}cm$ [13,20]. From measurements of spontaneous photon emissions in solids, which can be a consequence of the permanent energy increase, it was shown that this radius must be at least larger than $0.5\text{Å}$ [21]. Since this is larger than the spatial variation of the solid's nuclei, which will be calculated in section 3.2, the mass density distribution must be calculated in Diósi's model, different from that in Penrose's model. The differences resulting from a smearing of the mass-density operator with $R_{sm} \approx 10^{-5}cm$ will be outlined for all results in this paper, where the model without smearing of the mass-density operator will be referred to as the *parameter-free Diósi-Penrose model*.

# 3. Diósi-Penrose energy of superposed solids

In this section, we calculate the Diósi-Penrose energy of superposed solids that are slightly displaced with respect to each other or have slightly different expansions in the superposed states. The displacements shall always be significantly smaller than the dimensions of the solid, a restriction that is sufficient for most experiments. In section 3.1, we first simplify the calculation by neglecting the microscopic mass distribution resulting from the solid's nuclei by averaging the mass density. In section 3.2, we examine then the additional contribution resulting from the solid's nuclei, and discuss in section 3.3 the total Diósi-Penrose energy of superposed solids. In section 3.4 we look at the differences arising for the Diósi-Penrose model with smeared mass-density operator.

## 3.1. Contribution when spatially averaging the mass density

In this paper, we consider the four cases for superposed solids shown in figure 1. On the one hand, disks that are slightly displaced with respect to each other, or have slightly different extensions are considered, and on the other hand, a rod that has slightly different extensions in the superposed states is examined.

For the following calculation, we first assume that both the disk and the rod are infinitely extended. Under this assumption, the Diósi-Penrose energy can be calculated with equation (4). The gravitational field $g(\boldsymbol{x})$ is for the disk then always perpendicular to its plane and for the rod it points away from its centre. With the help of the law of flux, for the disk, we obtain $2g(x) = 4\pi G\rho 2x$, and for the rod, we obtain $2\pi r g(r) = 4\pi G\rho\pi r^2$, where $x$ is the distance from the centre of the disk, $r$ the distance from the centre of the rod, and $\rho$ the spatially averaged mass density of the solid. With $x' = x - \Delta s$ and $\rho' = \rho$ for the *displaced disk* (where the dashed variables refer to the second state), $x' = x - \Delta d/2$ and $\rho' = \rho(1 - \frac{\Delta d}{d})$ for the disk extended to the right, $x' = x$ and $\rho' = \rho(1 - \frac{\Delta d}{d})$ for the disk extended to both sides, and $r' = r$ and $\rho' = \rho(1 - 2\frac{\Delta r}{r})$ for the extended rod, we get for displacements, which are much smaller than the size of the solid ($\Delta s, \Delta d \ll d$, $\Delta r \ll r$), the following result:

$$E_{DP\,\overline{\rho}} = 2\pi G V \rho^2 \Delta s^2 \qquad \textit{displaced disk}$$
$$E_{DP\,\overline{\rho}} = \tfrac{1}{3}\, 2\pi G V \rho^2 \Delta d^2 \qquad \textit{disk extended to the right}$$
$$E_{DP\,\overline{\rho}} = \tfrac{1}{12}\, 2\pi G V \rho^2 \Delta d^2 \qquad \textit{disk extended to both sides} \qquad (6)$$
$$E_{DP\,\overline{\rho}} = \tfrac{1}{2}\, 2\pi G V \rho^2 \Delta r^2 \qquad \textit{extended rod}$$

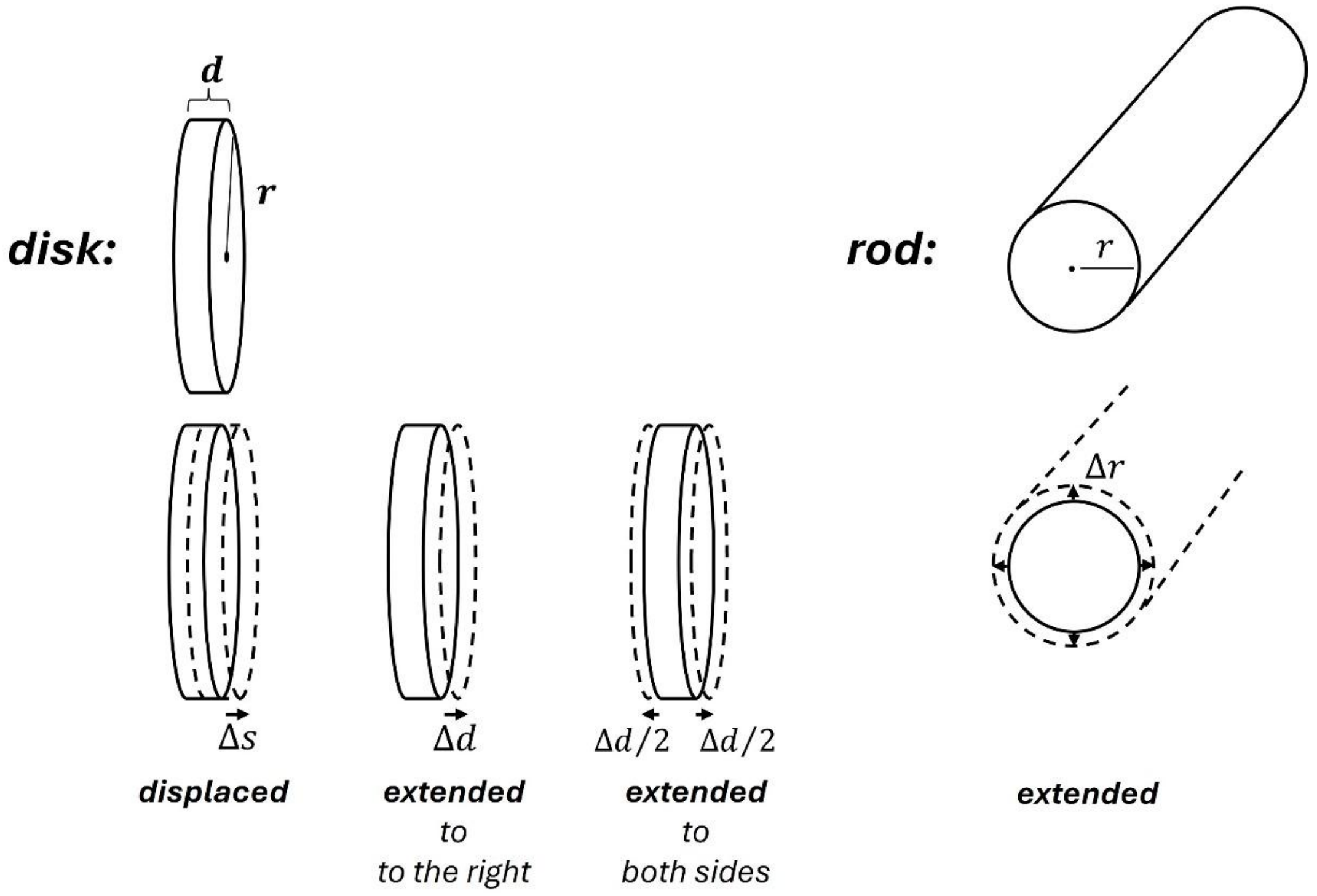

**Figure 1.** Discussed cases of solids in quantum superpositions. *Left:* Disk that is displaced by $\Delta s$ in one state relative to the other state or has slightly different thicknesses $\Delta d$ in the states (e.g. due to compression or piezoelectricity). *Right:* Rod that has slightly different radii $r$ in the states (e.g. due to thermal expansion).

***for:*** *disk and rod infinitely extended* ,

where $V$ is the volume of the solid. This result could suggest that the Diósi-Penrose energy of a superposed solid is generally proportional to its volume $V$, but which is not the case. For example, for a rod being displaced parallel to its direction, the Diósi-Penrose energy does not linearly increase with the rod's length. It converges to a constant value for large lengths of the rod.

In reference [1], the Diósi-Penrose energy was calculated for an oblate-like spheroid that is displaced perpendicular to its plane. For the case of small displacements $\Delta s$ and radii of the spheroid $r_{sph}$ that are much larger than its thickness $s$ ($\Delta s \ll d$, $r_{sph} \gg d$), the Diósi-Penrose energy is given by

$$E_{DP\,\overline{\rho}} = 2\pi G V \rho^2 \Delta s^2 \left(1 + \frac{\pi d}{4 r_{sph}}\right) \quad , \qquad (7)$$

where $V$ is the volume and $\rho$ the mass density of the spheroid. This result converges for $r_{sph} \gg d$, as expected, against the Diósi-Penrose energy of the *displaced disk* (6). If one assumes that the displaced spheroid leads to approximately the same Diósi-Penrose energy as the displaced disk if their thickness and volume are identical, which is the case for $r_{sph} = \sqrt{3/2}\, r$[1], we get by inserting this into (7) for the Diósi-Penrose energy of a disk with a finite radius $r$ the following correction to our results (6) for infinitely extended disks:

[1] $\frac{4\pi}{3} r_{sph}^2 \frac{d}{2} = \pi r^2 d$

$$
\begin{aligned}
&E_{DP\,\bar{\rho}} \approx 2\pi G V \rho^2 \Delta s^2 \left(1 + \tfrac{\pi}{\sqrt{24}} \tfrac{d}{r}\right) && \textit{displaced disk} \\
&E_{DP\,\bar{\rho}} \approx \tfrac{1}{3}\, 2\pi G V \rho^2 \Delta d^2 \left(1 + \tfrac{\pi}{\sqrt{24}} \tfrac{d}{r}\right) && \textit{disk extended to the right} \\
&E_{DP\,\bar{\rho}} \approx \tfrac{1}{12}\, 2\pi G V \rho^2 \Delta d^2 \left(1 + \tfrac{\pi}{\sqrt{24}} \tfrac{d}{r}\right) && \textit{disk extended to both sides} \quad .
\end{aligned}
\tag{8}
$$

### 3.2. Additional contribution due to the mass concentration in the nuclei

To understand how the Diósi-Penrose energy of the superposed solid changes, when one assumes that its mass is not uniformly distributed but concentrated in its nuclei, the following illustration of the Diósi-Penrose energy is helpful. Assuming hypothetically that the masses of the superposed state 1 and 2 attract each other by the gravitational force, the Diósi-Penrose energy describes the mechanical work to pull the masses of state 1 and 2 apart from each other against this gravitational attraction. This illustration of Diósi-Penrose energy only applies exactly when rigid objects are displaced against each other, as at the displaced disk in figure 1. The mechanical work to move the disk in state 2 over the distance $\Delta s$ in the gravitational potential of state 1 $\Phi_1(\mathbf{x})$ is given by $\int d^3\boldsymbol{x}\ (\rho_2(\mathbf{x}) - \rho_1(\mathbf{x}))\Phi_1(\mathbf{x})$ and is identical to the mechanical work to move the disk in state 1 in the gravitational potential of state $\Phi_2(\mathbf{x})$, which is given by $\int d^3\boldsymbol{x}\ (\rho_1(\mathbf{x}) - \rho_2(\mathbf{x}))\Phi_2(\mathbf{x})$. This directly leads to the Diósi-Penrose energy in the form of equation (3). From this illustration, it follows that for small displacements $\Delta s$, where the gravitational force can be linearised ($F_G \sim \Delta s$), the Diósi-Penrose energy increases quadratically with the displacement ($E_{DP} \sim \Delta s^2$), which agrees with our calculations in equation (6). The left part of figure 2 illustrates the gravitational forces $F_G$ between all partial masses of the solid in state 1 and 2 when the mass is spatially averaged. The right part of the figure shows the additional forces occurring when the mass of the solid is concentrated in its nuclei. For small displacements, neighbouring atomic nuclei attract each other strongly due to the high mass concentration in the nuclei. For larger displacement, the high mass concentration in the nuclei no longer plays a role, since the force between them does hardly change when the nuclei's masses are evenly distributed over the respective lattice cells. We therefore expect when taking the mass concentration in the nuclei into account that changes to our calculations with averaged mass density only occur for small displacements between the solids, smaller than a few lattice constants.

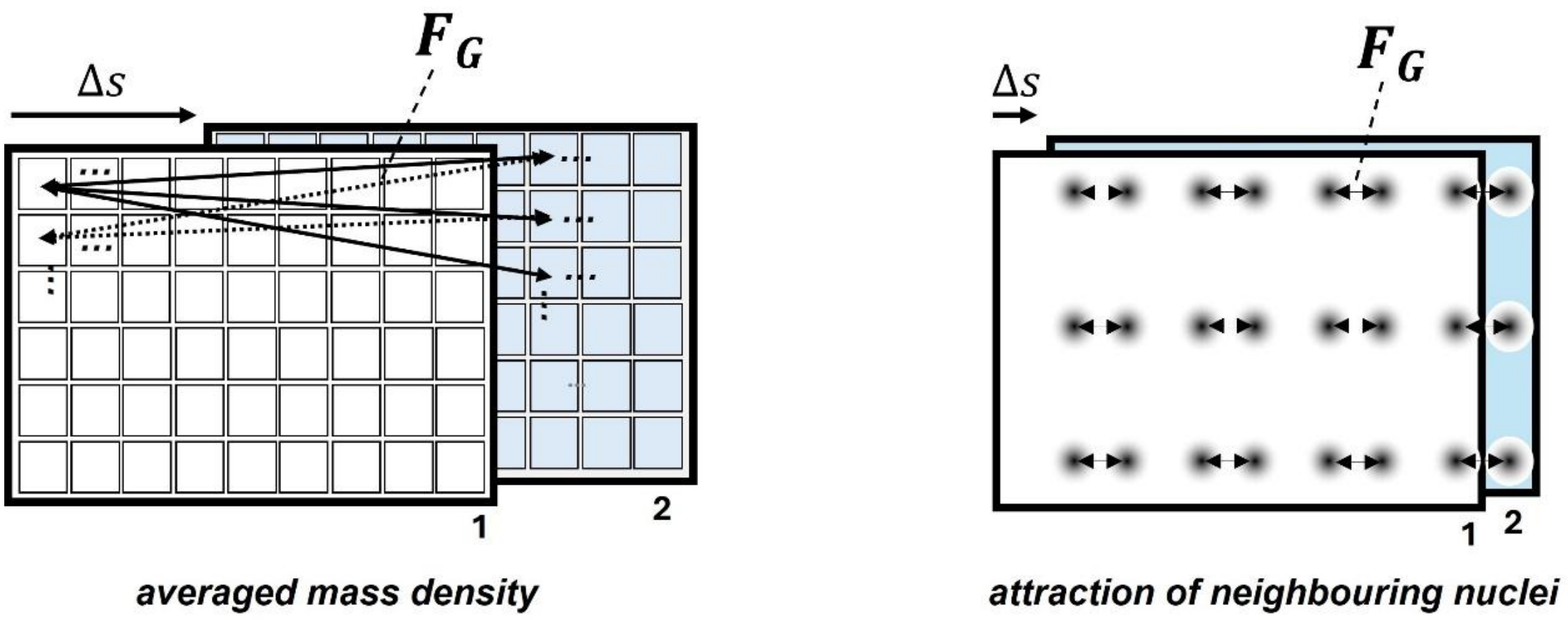

**Figure 2.** Illustration of the gravitational forces $F_G$ between the two states of a superposed solid displaced by $\Delta s$. *Left:* Forces when averaging the mass density of the solid. Here, each partial mass in state 1 attracts each partial mass in state 2, as illustrated in the figure. *Right*: Additional forces due to the mass concentration in the solid's nuclei. Here, strong forces occur between neighbouring nuclei if they are only slightly displaced against each other.

To calculate the additional contribution to the Diósi-Penrose energy due to the gravitational forces between neighbouring nuclei, we first calculate the Diósi-Penrose energy of a single superposed nucleus, where the mass distribution of the nucleus is assumed to be Gaussian as follows:

$$\rho(\boldsymbol{x}) = \frac{m}{\sqrt{2\pi}^3 \sigma^3} e^{-\frac{x^2}{2\sigma^2}} \quad . \tag{9}$$

Here $m$ is the nucleus' mass and $\sigma$ its spatial variation. The Diósi-Penrose energy of such a Gaussian mass distribution being displaced in state 1 and 2 by $\Delta s$ was calculated for the general case in reference [1] and for the limiting cases $x \ll 1$ and $x > 4$ in the appendix 2 of reference [22]. It leads to the following result:

$$E_{DP} = \frac{Gm^2}{\sqrt{\pi}\sigma} f_\sigma\left(\frac{\Delta s}{\sigma}\right) , \tag{10}$$

with

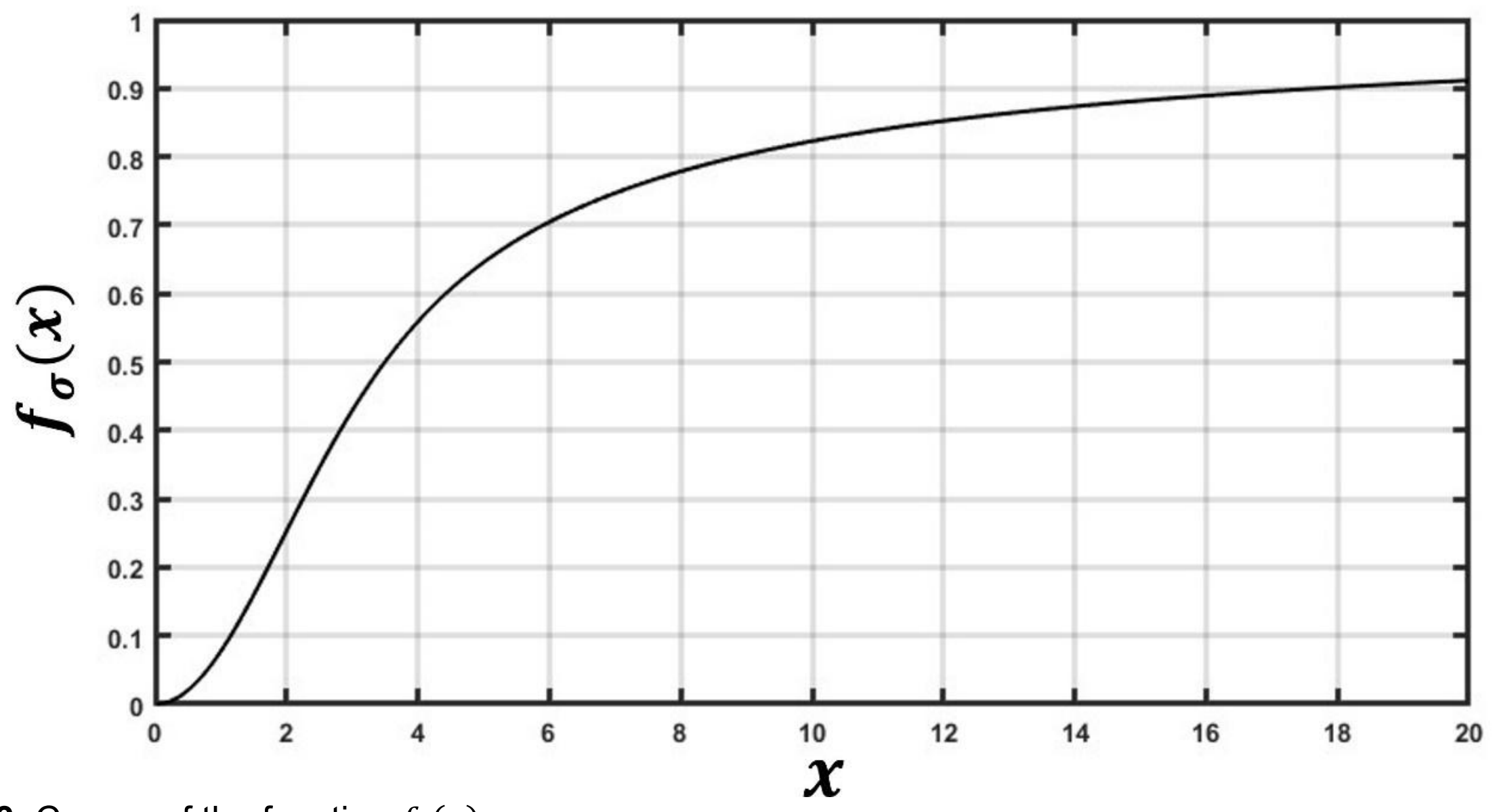

**Figure 3.** Course of the function $f_\sigma(x)$.

$$f_\sigma(x) = 1 - \frac{\sqrt{\pi}}{x} erf\left(\frac{x}{2}\right) \approx \begin{cases} \frac{1}{12}x^2 & x \ll 1 \\ 1 - \frac{\sqrt{\pi}}{x} & x > 4 \end{cases} . \tag{11}$$

The course of the function $f_\sigma(x)$ is shown in figure 3. For small displacement ($\Delta s \ll \sigma$), for which the gravitational force between the nuclei in the two states can be linearized ($F_G \sim \Delta s$), one obtains, as expected, a quadratic increase of the Diósi-Penrose energy over the displacement ($E_{DP} \sim \Delta s^2$), and for displacement $\Delta s > 4\sigma$, for which the nuclei in the two states are almost completely separated from each other and the gravitational force decreases with the inverse square of the distance ($F_G \sim 1/\Delta s^2$), the Diósi-Penrose energy converges to a constant value with $1/\Delta s$.

The spatial variation $\sigma$ of a nucleus in a solid is not of the order of the proton radius ($\approx 10^{-15} m$), but much larger, since the nucleus can oscillate around its rest position due to the excited phonons in the solid. The spatial variation $\sigma$ can be calculated with Debye's model for acoustical phonons and leads with the derivation in the appendix to:

$$\sigma = \sqrt{\frac{3\hbar}{\overline{m}\,\omega_D^3} \int_0^{\omega_D} d\omega\, \omega \left( \frac{1}{e^{\frac{\hbar\omega}{k_B T}} - 1} + \frac{1}{2} \right)} , \tag{12}$$

where $\omega_D$ is the Debye frequency and $\overline{m}$ the mean atomic mass of the solid, which is defined by:

$$\overline{m} \equiv \sum_i \frac{N_i}{N} m_i . \tag{13}$$

Here $N$ is the total number of atoms and $N_i$ the number of atoms with mass $m_i$. The Debye frequency $\omega_D$ can either be calculated from the solid's Debye temperature $\Theta_D$ by

$$\hbar\, \omega_D = k_B\, \Theta_D , \tag{14}$$

which is determined from the temperature profile of the solid's specific heat, or from the sound velocities of the solid, as shown in the appendix. Table 1 shows the spatial variation of the nuclei $\sigma$ calculated with equations (12) - (14) for selected solids from the parameters on the left for zero and room temperature. These spatial variations are at room temperature typically in the order of a tenth of an Ångström and about a factor of two smaller at zero temperature. The calculated value for copper at zero temperature in table 1 of $\sigma = 0.043$Å agrees precisely with the value reported in reference [20], which was determined from a measurement of the Debye-Waller factor.

To calculate the total Diósi-Penrose energy of a displaced solid, the Diósi-Penrose energies of all nuclei of the solid must be added (see equation (10)):

$$E_{DP\,nucl} = \sum_{nucl\ i} \frac{G m_i^2}{\sqrt{\pi}\sigma} f_\sigma\left(\frac{\Delta s_i}{\sigma}\right) \ . \tag{15}$$

In this formula the spatial variations $\sigma$ of all nuclei are assumed to be the same, even if they have different masses as in solids consisting of several chemical elements. This assumption is justified by the fact that in Debye's model the acoustical phonons are relevant for the spatial variation of the nuclei, in which neighbouring atoms oscillate approximately in phase. In approach (15), we assume amorphous solids. In perfect single crystals and displacements by exactly one lattice constant along a lattice axis, there occur strong attractive forces to the neighbouring nuclei of the other state, and the mechanical work to displace the states against each other by one lattice constant would then be zero.

Approach (15) leads for the displaced disk, where all neighbouring atomic nuclei are displaced the same distance $\Delta s$, to:

$$E_{DP\,nucl} = T_{DP}^{solid}\, V\ f_\sigma\left(\tfrac{\Delta s}{\sigma}\right) \qquad \textit{displaced disk}\ , \tag{16}$$

where the quantity $T_{DP}^{solid}$ is defined by:

$$T_{DP}^{solid} = \frac{G}{\sqrt{\pi}} \frac{\overline{m^2}}{\bar{m}^2} \frac{\bar{g}^3}{\sigma}\ \rho^2 \ . \tag{17}$$

Here, $\bar{g}$ is the mean lattice constant of the solid, which can be calculated from the mass density $\rho$ and the mean atomic mass $\bar{m}$ (equation (13)) of the solid by:

$$\bar{g} = \sqrt[3]{\frac{\bar{m}}{\rho}} . \tag{18}$$

The factor $\overline{m^2}/\bar{m}^2$ in (17) is called the *square mass factor* of the solid, which is given by:

$$\frac{\overline{m^2}}{\bar{m}^2} \equiv \frac{\sum_i \frac{N_i}{N} m_i^2}{\bar{m}^2} \quad . \tag{19}$$

For solids consisting of one chemical element the square mass factor is one. For solids consisting of more than one element the term $m_i^2$ in (19) considers that the Diósi-Penrose energy of a single nucleus is according to equation (10) proportional to the square of its mass.

When calculating the other three cases in figure 1, it must be taken into account that the distances between neighbouring nuclei depends then on the location inside the solid, which leads to the following modification of our result (16):

$$E_{DP\,nucl} = T_{DP}^{solid}\,V\;\frac{\int_0^{\frac{\Delta d}{\sigma}} dx\;\; f_\sigma(x)}{\frac{\Delta d}{\sigma}} \qquad \textit{disk extended to the right}$$

$$E_{DP\,nucl} = T_{DP}^{solid}\,V\;\frac{\int_0^{\frac{\Delta d}{2\sigma}} dx\;\; f_\sigma(x)}{\frac{\Delta d}{2\sigma}} \qquad \textit{disk extended to both sides} \qquad (20)$$

$$E_{DP\,nucl} = T_{DP}^{solid}\,V\;\frac{\int_0^{\frac{\Delta r}{\sigma}} 2\pi r dr\;\; f_\sigma(r)}{\pi\left(\frac{\Delta r}{\sigma}\right)^2} \qquad \textit{extended rod} \quad ,$$

For displacements being much larger than the spatial variation of the nuclei, equations (16) and (20) lead for all four cases in figure 1 to the same result[2]:

$$E_{DP\,nucl} = T_{DP}^{solid}\,V \qquad \textbf{\textit{for:}}\;\; \Delta s, \Delta d, \Delta r \gg \sigma \;\; . \qquad (21)$$

The term $T_{DP}^{solid}\,V$ describes according to the illustration in figure 2 the energy required to pull the nuclei of the solid in the two states far apart from each other against their gravitational attraction. The quantity $T_{DP}^{solid}$ we denoted as the *characteristic Diósi-Penrose energy density* of the solid. The characteristic Diósi-Penrose energy divided by Planck's constant $T_{DP}^{solid}/\hbar$ describes a decay rate per volume (see equation (1)). Table 1 shows $T_{DP}^{solid}/\hbar$ for the selected solids, which are calculated from the parameters on the left in the table with equation (17). $T_{DP}^{solid}/\hbar$ ranges from $3.7\,MHz/cm^3$ for silicon dioxide to $34.1\,MHz/cm^3$ for iron and $370.2\,MHz/cm^3$ for platinum at room temperature and is about two times larger at zero temperature.

For displacements much smaller than the spatial variation of the nuclei, where the Diósi-Penrose energy increases quadratically with the displacement, equations (16) and (20) lead with $f_\sigma(x) = \frac{1}{12}x^2$ for $x \ll 1$ to:

$$E_{DP\,nucl} = T_{DP}^{solid}\,V\frac{1}{12}\left(\frac{\Delta s}{\sigma}\right)^2 \qquad \textit{displaced disk}$$

$$E_{DP\,nucl} = \frac{1}{3}\,T_{DP}^{solid}\,V\frac{1}{12}\left(\frac{\Delta d}{\sigma}\right)^2 \qquad \textit{disk extended to the right}$$

$$E_{DP\,nucl} = \frac{1}{12}\,T_{DP}^{solid}\,V\frac{1}{12}\left(\frac{\Delta d}{\sigma}\right)^2 \qquad \textit{disk extended to both sides} \qquad (22)$$

$$E_{DP\,nucl} = \frac{1}{2}\,T_{DP}^{solid}\,V\frac{1}{12}\left(\frac{\Delta r}{\sigma}\right)^2 \qquad \textit{extended rod}$$

$$\textbf{\textit{for:}}\;\; \Delta s, \Delta d, \Delta r \ll \sigma \quad .$$

It is interesting to mention that we get here for the cases in figure 1 the same pre-factors as calculated for solids with averaged mass density in equation (6), namely 1 (displaced disk), $\frac{1}{3}$ (disk extended to the right), $\frac{1}{12}$ (disk extended to both sides ) and $\frac{1}{2}$ (extended rod).

[2] Note that $f_\sigma(x) = 1$ for $x \gg 1$.

| | $\rho\left[\frac{g}{cm^3}\right]$ | $\bar{m}[u]$ | $\frac{\overline{m^2}}{\bar{m}^2}$ | $\bar{g}[Å]$ | $\Theta_D[K]$ | $\sigma[Å]$ (T = 0 K) | $\frac{T_{DP}^{solid}}{\hbar}\left[\frac{MHz}{cm^3}\right]$ (T = 0 K) | $\xi$ (T = 0 K) | $\sigma[Å]$ (T = 300 K) | $\frac{T_{DP}^{solid}}{\hbar}\left[\frac{MHz}{cm^3}\right]$ (T = 300 K) | $\xi$ (T = 300 K) |
|---|---|---|---|---|---|---|---|---|---|---|---|
| $Al$ | 2.70 | 26.98 | 1 | 2.55 | 390 | 0.059 | 7.3 | 611 | 0.105 | 4.1 | 106 |
| $Si$ | 2.34 | 28.09 | 1 | 2.71 | 692 | 0.043 | 9.0 | 1844 | 0.061 | 6.4 | 663 |
| $Fe$ | 7.87 | 55.84 | 1 | 2.28 | 373 | 0.042 | 62.4 | 1207 | 0.077 | 34.1 | 197 |
| $Cu$ | 8.96 | 63.55 | 1 | 2.28 | 310 | 0.043 | 78.5 | 1110 | 0.086 | 39.3 | 140 |
| $Pb$ | 11.34 | 207.2 | 1 | 3.12 | 87 | 0.045 | 310.1 | 2504 | 0.167 | 83.4 | 49 |
| $Au$ | 19.32 | 197.0 | 1 | 2.57 | 178 | 0.032 | 700.3 | 3790 | 0.084 | 268.4 | 213 |
| $Pt$ | 21.45 | 195.1 | 1 | 2.47 | 225 | 0.029 | 861.6 | 4736 | 0.067 | 370.2 | 376 |
| $SiO_2$ | 2.65 | 20.03 | 1.08 | 2.32 | 523 | 0.059 | 5.8 | 496 | 0.093 | 3.7 | 127 |
| $Al_2O_3$ | 4 | 20.39 | 1.07 | 2.04 | 980 | 0.043 | 12.1 | 872 | 0.053 | 9.7 | 450 |
| $PZT$[3] | 7.6 | 64.95 | 2.32 | 2.42 | 266 | 0.046 | 148.1 | 2553 | 0.099 | 69.0 | 258 |

**Table 1.** Spatial variation of nuclei $\sigma$, characteristic Diósi-Penrose energy density $T_{DP}^{solid}$ and the factor $\xi$, describing how much the Diósi-Penrose energy increases due to the mass concentration in the nuclei for small displacements (see section 3.3) for zero and room temperature calculated for selected solids from their material properties on the left.

### 3.3. Both contributions

In this section we consider both contributions to the Diósi-Penrose energy, the contribution of the nuclei $E_{DP\,nucl}$ and the contribution when spatially averaging the mass density of the solid $E_{DP\,\bar{\rho}}$, and investigate which contribution dominates in which range of displacement. The total Diósi-Penrose energy of a superposed solid is given by the sum of these two contributions:

$$E_{DP} = E_{DP\,\bar{\rho}} + E_{DP\,nucl} \quad , \tag{23}$$

where $E_{DP\,\bar{\rho}}$ is given by (6) or (8) and $E_{DP\,nucl}$ by (16) or (20) for our four cases in figure 1.

For displacement being much smaller than the spatial variation of the nuclei ($\Delta s \ll \sigma$), the contribution of the nuclei $E_{DP\,nucl}$ is by the factor

$$\frac{E_{DP\,nucl}}{E_{DP\,\bar{\rho}}} \equiv \xi = \frac{1}{24\pi^{3/2}} \left(\frac{\overline{m^2}}{\bar{m}^2}\right) \left(\frac{\bar{g}}{\sigma}\right)^3 \qquad \textbf{\textit{for: }} \Delta s, \Delta d, \Delta r \ll \sigma \tag{24}$$

larger than the contribution when spatially averaging the mass density $E_{DP\,\bar{\rho}}$, whereby this factor is the same for all cases in figure 1[4]. Table 1 shows this factor $\xi$ for the selected solids. It ranges from $50$ up to $650$ at room temperature and is at zero temperature larger ($500$ - $5000$). Since the contribution of the nuclei $E_{DP\,nucl}$ converges for large displacements towards the constant value $T_{DP}^{solid}\,V$ (equation (21)), the quadratically increasing contribution $E_{DP\,\bar{\rho}}$ (equations (6)) dominates the Diósi-Penrose energy above a certain displacement $\Delta s$. For the displaced disk, $E_{DP\,\bar{\rho}}$ reaches $T_{DP}^{solid}\,V$ for a displacement $\Delta s$ of:

[3] $Pb(Zr_xTi_{1-x})O_3$ with $x \approx 0.5$.

[4] This follows by dividing equations (22) by equations (6).

$$\left.\frac{\Delta s}{\bar{g}}\right|_{E_{DP\,\bar{\rho}} = T_{DP}^{solid}V} = \sqrt{\frac{1}{2\pi^{3/2}}\left(\frac{\overline{m^2}}{\overline{m}^2}\right)\frac{\bar{g}}{\sigma}} \qquad \textit{displaced disk} \quad . \tag{25}$$

For the solids in table 1, $\Delta s/\bar{g}$ ranges from $1.5$ to $2.2$ at room temperature and from $1.9$ to $3.2$ at zero temperature. This means that for displacements larger than of about ten lattice constants, the contribution of the nuclei $E_{DP\,nucl}$ to the Diósi-Penrose energy of the solid can be neglected in a good approximation:

$$E_{DP} \approx E_{DP\,\bar{\rho}} \quad . \qquad \textbf{\textit{for:}}\ \Delta s, \Delta d, \Delta r > 10\bar{g} \tag{26}$$

Figure 4 shows the typical course of the Diósi-Penrose energy of a solid over the displacement $\Delta s$ (black curve) and of its two contributions $E_{DP\,\bar{\rho}}$ (blue curve) and $E_{DP\,nucl}$ (dashed curve) calculated for a solid with the typical ratio between the mean lattice constant and the spatial variation of its nuclei at room temperature of $\bar{g}/\sigma = 30$ (see values in table 1) and $\overline{m^2}/\overline{m}^2 = 1$ and for the case of a displaced disk. This illustration clearly shows that the contribution of the nuclei $E_{DP\,nucl}$ dominates the Diósi-Penrose energy for small displacements and can be neglected for large ones.

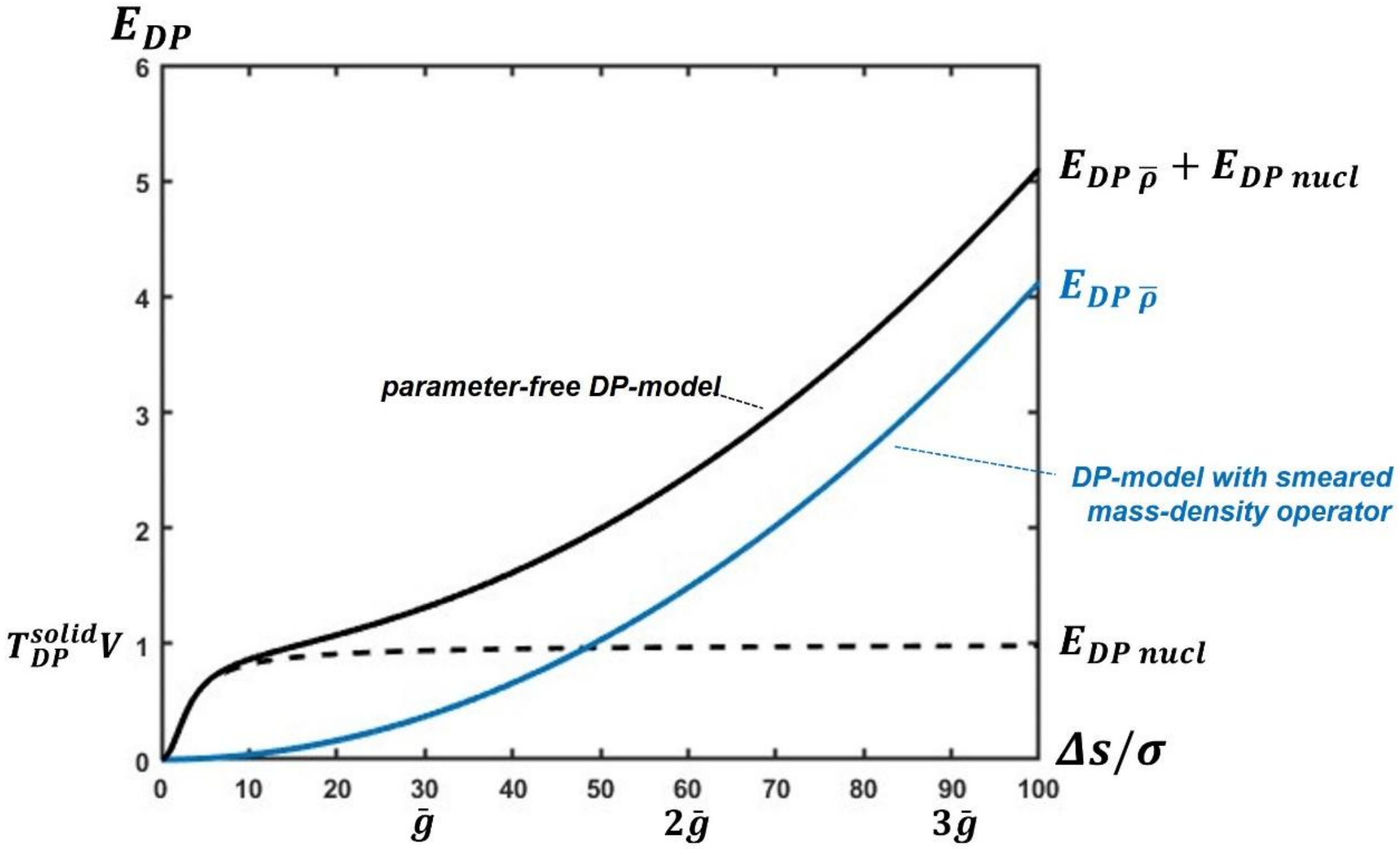

**Figure 4.** Typical course of the Diósi-Penrose energy of a solid over the displacement $\Delta s$ (black curve) and of its contributions $E_{DP\,\bar{\rho}}$ (blue curve) and $E_{DP\,nucl}$ (dashed curve) calculated for a solid with $\bar{g}/\sigma = 30$ and $\overline{m^2}/\overline{m}^2 = 1$ and the case of a displaced disk.

### 3.4. Model with smeared mass-density operator

In this section, we examine how the derived results change when regarding the *Diósi-Penrose model with smeared mass-density operator* in section 2.1. Since the proposed radius for smearing the mass density of $R_{sm} = 10^{-5} cm$ (section 2.1) is about $500$ times larger than the mean lattice constant $\bar{g}$ of the solids (see table 1), the averaging procedure leads to an almost homogeneous mass distribution, and the Diósi-Penrose energy can be calculated as done in section 3.1 by spatially averaging the mass density and the contribution of the nuclei can be neglected. This leads to:

$$E_{DP} = E_{DP\,\bar{\rho}} \quad . \qquad \textbf{for: } R_{sm} \gg \bar{g} \tag{27}$$

From the discussion in section 3.3 it follows that for displacements smaller than the spatial variation of the nuclei the Diósi-Penrose model with smeared mass-density operator leads to significant smaller decay rates than the parameter-free model, which are by the factor $\xi$ in table 1 smaller, and that for displacements larger than ten lattice constants both models hardly differ. In figure 4, the course of the Diósi-Penrose energy over the displacement for the Diósi-Penrose model with smeared mass-density operator is highlighted as a blue curve.

# 4. Application to electrical components

In this section, we apply our results for superposed solids of section 3 to electrical components. In sections 4.1 and 4.2 we regard components that are needed for the construction of indirect single-photon detectors, which are plate capacitors and current-carrying components such as resistors, wires and photodiodes. In section 4.3 we discuss piezo-actuators, which can be used in conjunction with indirect detectors to transfer solids, as e.g. a mirror, into a quantum superposition.
The electrical components discussed in in this section evolve into quantum superpositions when they are part an indirect single-photon detector. As we will show in section 5, the indirect single-photon detector stays for some time in a superposition of a photon-detected and a no-photon-detected state, and this also applies to its electrical components. For example, in a resistor of the detector a current may has flown in the photon-detected state but not in the no-photon-detected state, which leads to a warming and thus slight expansion of the resistor in the photon-detected state. From this we can calculate with our results for solids a Diósi-Penrose energy for the superposed resistor.

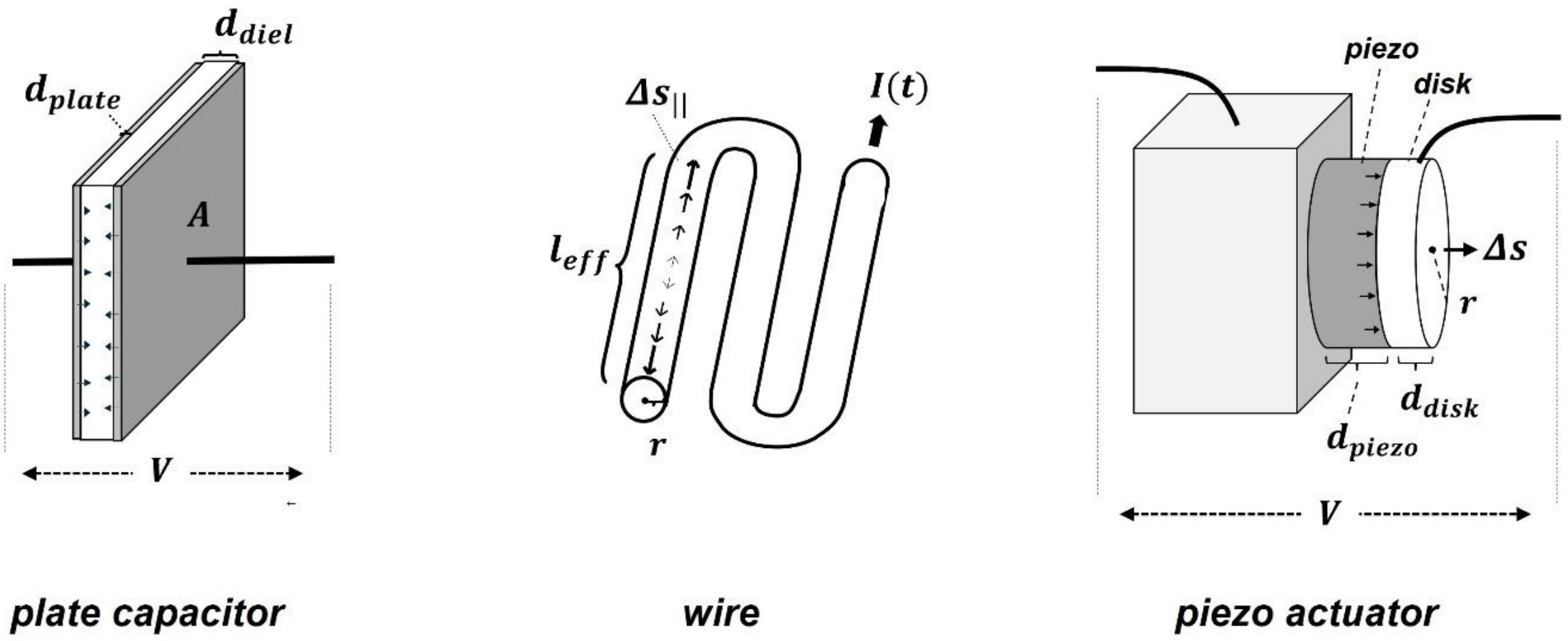

**Figure 5.** Electrical components discussed in sections 4.1, 4.2 and 4.3.

### 4.1. Plate capacitor

In this section, we consider the plate capacitor in the left of figure 5, to which different voltages $V$ are applied in the two superposed states and consequently the capacitor is charged to different degrees in these states. The main effect that leads to different mass distributions in these states and is therefore relevant for the Diósi-Penrose energy of the superposed capacitor is the compression of its dielectric caused by the electric attraction of its plates. Other effects as the polarization of the dielectric and the different mass densities on the capacitor plates resulting from the charged electrons are orders of magnitude smaller as shown in reference [22].

The compression of the capacitor's dielectric of thickness $d_{diel}$ by $\Delta d_{diel}$ due to a change of pressure $\Delta P$ of the capacitor's plates on the dielectric can be calculated with the modulus of elasticity of the dielectric $E_{el}$ by:

$$\frac{\Delta d_{diel}}{d_{diel}} = -\frac{\Delta P}{E_{el}} \quad . \tag{28}$$

When the voltage $V$ is in one state of the capacitor by $\Delta V$ higher than in the other, the dielectric will be compressed in this state by: [5]

$$\Delta d_{diel} = \frac{2\varepsilon_0 \varepsilon_r^2 V \Delta V}{E_{el}\, d_{diel}} \quad , \tag{29}$$

where $\varepsilon_0$ is the electric constant and $\varepsilon_r$ the dielectric's relative permittivity.

Since this compression of the dielectric is, as the calculations in section 5.2 will show, usually significantly smaller than the spatial variation of the nuclei, the Diósi-Penrose energy of the superposed capacitor is determined for the parameter-free Diósi-

[5] This follows with relations $P = F/A$, $F = QE$, $\varepsilon_0 AE = Q$, $C = Q/V$ and $C = \varepsilon_0 \varepsilon_r A/d$, where $F$ is the force between the plates, $E$ the electric field outside the dielectric (which is relevant to calculate the force between the plates), $Q$ the charge on the plates, and $C$ the capacitance.

Penrose model by the contribution of the nuclei only. The Diósi-Penrose energy can then be calculated with equations (22) for the disk extended to both sides and the displaced disk as[6]:

$$E_{DP} \approx \frac{A}{12}\left(\frac{d_{diel}\, T_{DP}^{diel}}{12\, \sigma_{diel}^{2}} + \frac{d_{plate}\, T_{DP}^{plate}}{2\, \sigma_{plate}^{2}}\right) \Delta d_{diel}{}^{2} \quad . \qquad \textit{parameter-free DP-model} \qquad (30)$$

**for:** $\Delta d_{diel} \ll \sigma$

Here $A$ is the area of the capacitor, $d_{diel}$ and $d_{plate}$ the thicknesses for the capacitor's dielectric and the plates (see figure 5), and $\sigma_{diel}$, $\sigma_{plate}$, $T_{DP}^{diel}$ and $T_{DP}^{plate}$ their spatial variations of nuclei and characteristic Diósi-Penrose energy densities (17).

Since the contribution of the nuclei is not relevant for the Diósi-Penrose model with smeared math-density operator, the Diósi-Penrose energy of the capacitor must be calculated with equations (6) instead of (22), which leads to:

$$E_{DP} \approx 2\pi G A \left(\frac{1}{12} d_{diel}\, \rho_{diel}^{2} + \frac{1}{2} d_{plate} \rho_{plate}^{2}\right) \Delta d_{diel}{}^{2} \;, \qquad \textit{DP-model with smeared mass-density operator} \qquad (31)$$

where $\rho_{diel}$ and $\rho_{plate}$ are the mass densities of the dielectric and the plates.

In our results (30) and (31) we simply added the Diósi-Penrose energies resulting from the dielectric and from the two plates of the capacitor together. This is exactly valid for the contribution of the nuclei, i.e. for result (30), but for result (31) only, when the plates of the capacitor are infinitely extended. This can be verified with equation (4) as follows. From the law of flux it follows that the difference in the gravitational fields $g_1(\boldsymbol{x}) - g_2(\boldsymbol{x})$ belonging to the mass distributions $\rho_1(\mathbf{x})$ and $\rho_2(\mathbf{x})$ in the two states of the superposed dielectric or of the two plates are exactly zero outside the respective component, when they are infinitely extended. Then, when calculating the Diósi-Penrose energy of the plate capacitor with equation (4), the contributions of the dielectric and of the two plates do not interfere and can simply be added.

## 4.2. Current-carrying components

In this section, we discuss current-carrying components such as resistors and wires. When these components are part of a photon detector, they can differ in the photon-detected state and no-photon-detected state that in the photon-detected state a current $I(t)$ has flowed through them, and in the other state not. A current flow $I(t)$ leads to the following temperature increase of the component:

$$\Delta T = \frac{R}{cM}\int_0^{t_s} dt I(t)^2 \qquad , \qquad (32)$$

where $R$ is the resistance and $M$ the mass of the component, and $c$ the specific heat of the material. If we consider the wire in the middle of figure 5 exemplary for a current-carrying component, an expansion of the wire leads to displacements of its nuclei at the surface by $\Delta r$, but also to displacements $\Delta s_{||}$ along the wire as illustrated in the figure. For the contribution $E_{DP\,\bar{\rho}}$, when spatially averaging the mass density (section 3.1), only the extension of the rod perpendicular to its surface by $\Delta r$ is relevant. But, for the contribution of the nuclei $E_{DP\,nucl}$ the displacements along the wire $\Delta s_{||}$ are much larger than the perpendicular ones when the length of the wire is much larger than its diameter. For the calculation of $\Delta s_{||}$ only the straight sections of the wire without bending can be used, where the mean length of these straight sections is denoted as effective length $l_{eff}$ as illustrated in figure 5. The change of the wire's radius $\Delta r$ and of its effective length $\Delta l_{eff}$ due to a temperature increase $\Delta T$ can be calculated with the thermal expansion coefficient of the material $\alpha$ as follows:

$$\begin{aligned} \Delta l_{eff} &= \alpha\ l_{eff}\,\Delta T \\ \Delta r &= \alpha\ r\,\Delta T \end{aligned} \quad . \tag{33}$$

[6] Note that the plates are displaced by $\Delta d_{diel}/2$.

As for the plate capacitor, the calculations in section 5.2 will show that these displacements are much smaller than the spatial variation of the nuclei ($\Delta l_{eff}, \Delta r \ll \sigma$), we must consider for the parameter-free Diósi-Penrose model only the contribution of the nuclei. Here the effect of the displacements of the nuclei parallel to the wire $\Delta s_{||}$ are relevant, which must be calculated with the formula for the disk extended to both sides in equation (22), which leads to:

$$E_{DP} \approx \frac{1}{144} T_{DP}^{solid}\, V \left(\frac{\Delta l_{eff}}{\sigma}\right)^2 \quad . \tag{34}$$

*parameter-free DP-model*

***for:*** $\Delta l_{eff} \ll \sigma$

For the Diósi-Penrose model with smeared math-density operator we get from the extension of the wire's diameter by $\Delta r$ with equation (6) for the extended rod the following result:

$$E_{DP} \approx \pi G V \rho^2 \Delta r^2 \quad . \tag{35}$$

*DP-model with smeared mass-density operator*

## 4.3. Piezo-actuator

The right part of figure 5 shows a piezo-actuator displacing a disk to the right by $\Delta s$ when a voltage $V$ is applied to it. We regard a piezo for which only the $d_{33}$-component of the matrix for the converse piezoelectric effect is relevant, where $d_{33}$ describes the extension of the piezo in z-direction when an electrical field $E$ in the same direction is applied ($\Delta d/d = d_{33}E$). Applying a voltage $V$ to the piezo-actuator leads then to the following displacement of the disk:

$$\Delta s = d_{33} V \quad . \tag{36}$$

The calculations in section 5.3 for a single-photon detector steering such a piezo-actuator yield displacements $\Delta s$ being larger than ten lattice constants of the solids ($\Delta s > 10\bar{g}$). According to the discussion in sections 3.3 and 3.4 the contribution of the nuclei can be neglected when calculating the Diósi-Penrose energy and the parameter-free and the Diósi-Penrose model with smeared mass-density operator lead to approximately the same results. With result (8) for the disk extended to the right and the displaced disk we get for the piezo-actuator consisting of the piezo and the disk the following Diósi-Penrose energy:

$$E_{DP} \approx 2\pi G \left(\tfrac{1}{3}\rho_{piezo}^2 d_{piezo} \left(1 + \tfrac{\pi}{\sqrt{24}} \tfrac{d_{piezo}}{r}\right) + \rho_{disk}^2 d_{disk} \left(1 + \tfrac{\pi}{\sqrt{24}} \tfrac{d_{disk}}{r}\right)\right) \pi r^2 \Delta s^2 \,, \qquad (37)$$

$$\textbf{for: } \Delta s > 10\bar{g}$$

where $r$ is the radius of the disk, $d_{piezo}$ and $d_{disk}$ the thicknesses of the piezo and the disk, and $\rho_{piezo}$ and $\rho_{disk}$ their mass densities.

# 5. Application to indirect single-photon detectors

In this section, we apply our results derived in section 4 for electrical components to indirect single-photon detectors and investigate how long they stay in a superposition of a photon-detected and a no-photon-detected state, whereby all calculations will refer to room temperature. This choice may surprise the reader, since decoherence between superposed states due to interactions with the environment, such as electromagnetic radiation, vibrations and thermal couplings, strongly increases with temperature, and experiments to measure wavefunction collapse are therefore carried out at very low temperatures [2,4,6,7]. However, as mentioned in the introduction (section 1), this work focuses on the calculation of the lifetimes of superpositions and not on decoherence times. Since decoherence times decrease strongly with increasing temperature, they are expected to be much shorter at room temperature than the lifetimes of the superpositions calculated in the following. To clearly distinguish between the phenomenon of decoherence and wavefunction collapse, physical effects that lead to decoherence are only of interest here when they give rise to different mass-density distributions in the superposed states and shorten the lifetime of the superposition according to the Diósi–Penrose criterion. The discussion in section 5.2 of all effects that give rise to different mass-density distributions and thus determine the lifetime of the superpositions will show that physical effects leading to decoherence, such as electromagnetic radiation, vibrations and thermal couplings, can be neglected.

In section 5.1 we introduce the indirect single-photon detectors and calculate in section 5.2 how long their components stay in a superposition during detection. In section 5.3 we show how an indirect detector, combined with a piezo-actuator, can be used to transfer a mirror into a quantum superposition of slightly different positions, and in section 5.4 we show how the lifetime of indirect detectors can be deliberately controlled with a piezo-actuator displacing a mass.

## 5.1. Indirect single-photon detectors

In this section, we show how indirect single-photon detectors can be realized that interact as little as possible with their environment during detection and whose measurement outcome (whether a photon has been detected) is communicated to the outside after the superposition of the photon-detected and no-photon-detected states has already collapsed. Indirect detectors are therefore be characterized by "measurement after reduction" rather than "reduction by measurement". Here, we restrict ourselves to the basic concept of indirect single-photon detectors and leave the technical implementation out of scope, such as how the electronic switches in figure 6 used to decouple the detector from the environment can be realized, or how, in a specific experiment, it can be ensured that the detector has already been decoupled from the environment at the time the photon arrives.

Figure 6 shows the proposed setup for an indirect single-photon detector using a single-photon avalanche photodiode (SPAD) for the photon detection, which is biased above its breakdown voltage $V_B$ by an excess bias voltage $V_E$. The arrival of a single photon can then trigger with a certain probability (the quantum efficiency) an avalanche current and leads to a diode breakdown. To avoid any interaction with the power grid during detection, the photodiode is biased by a simple plate capacitor rather than a conventional voltage source. This capacitor is charged shortly before the photon's arrival using a conventional voltage source, which is then disconnected from the capacitor prior to detection by opening the two upper switches in figure 6. The measurement result is obtained after a sufficient time has passed for the superposed states in the detector to collapse, by connecting a voltmeter to the plate capacitor via the two lower switches in figure 6 and measuring whether the capacitor's voltage has dropped due to an avalanche current caused by photon detection. The four switches in figure 6 used to disconnect the voltage source and connect the voltmeter could be realized using field-effect transistors.

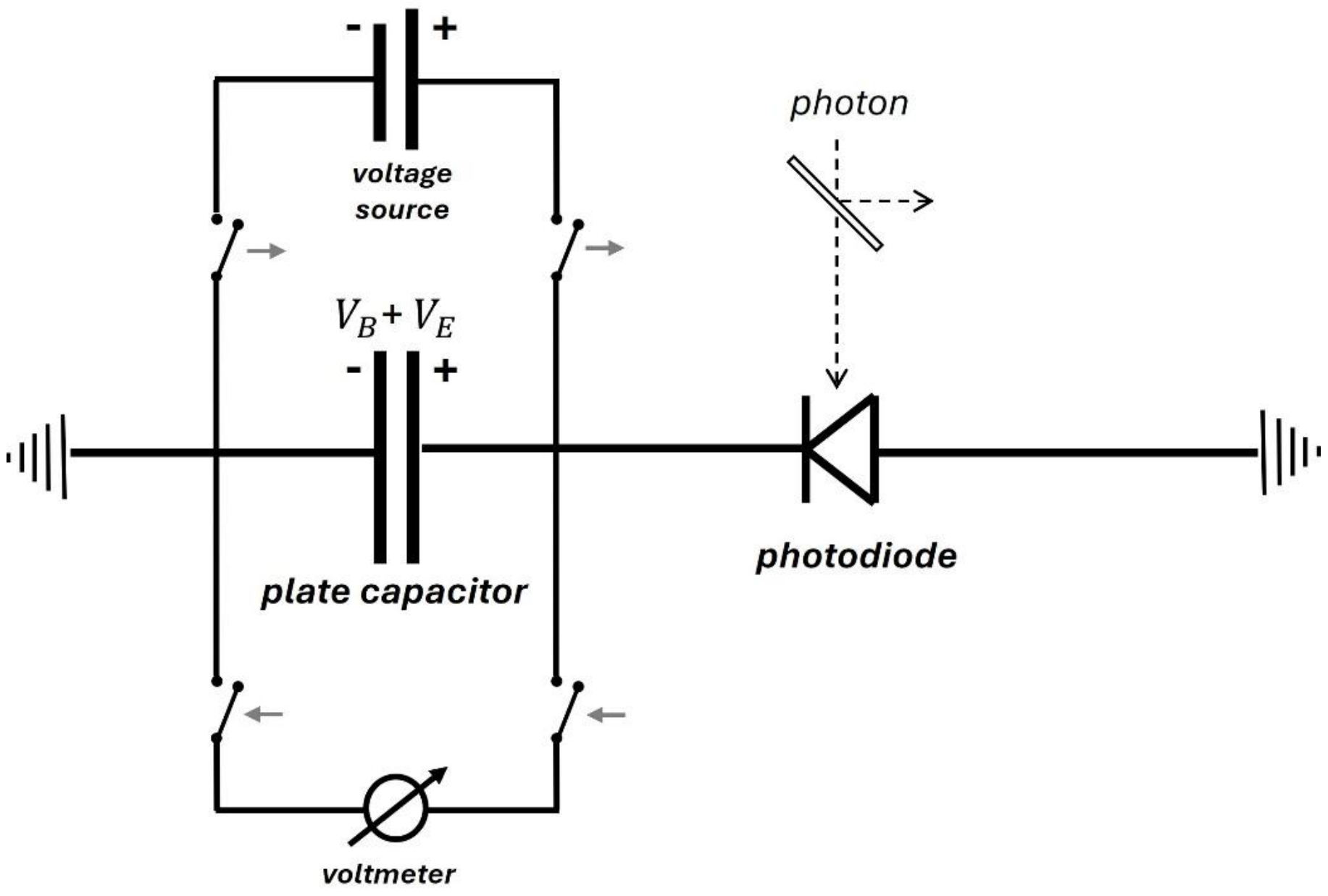

**Figure 6.** Indirect single-photon detector interacting as little as possible with its environment during detection, and whose detection result is read out after the detector has already undergone reduction (measurement after reduction).
To avoid interaction with an external voltage source during detection, the photodiode is biased by a plate capacitor that is charged by the voltage source prior to detection and disconnected from it during detection by opening the two upper switches. The measurement result is determined a sufficiently long time after the superposition of the photon-detected and no-photon-detected states has collapsed. This is done by connecting a voltmeter to the plate capacitor by closing the two lower switches and measuring whether the capacitor voltage has dropped due to an avalanche current triggered by a photon detection.

At this point, it is important to note that the assumption that the indirect single-photon detector can be completely isolated from its environment during detection by means of the four switches in figure 6 is idealized. In practice, there will always be interactions with the environment, for example due to leakage currents or thermal coupling. According to the Diósi–Penrose criterion, these interactions, although they may contribute significantly to the decoherence of the superposed states, affect the lifetime of the superposition only if they lead to different mass density distributions in the states. The discussion in the following chapter of all effects that give rise to different mass density distributions in the indirect single-photon detector will show that the interactions of the detector with its environment can be neglected compared to the effects inside the detector.

Before calculating the lifetime of the superposition of the photon-detected and no-photon-detected states in the following, we briefly discuss how such superpositions arise in the detector. In most experiments, this occurs because the photon spreads over a spatial region prior to measurement or is split by a beam splitter, as shown in figure 6, leading to a finite probability to detect the photon at the location of the detector and thus to a superposition of a photon-detected and a no-photon-detected state. However, even when a localized photon is measured directly, the detector is still expected to evolve into a superposition of a photon-detected and a no-photon-detected state due to its finite detection probability (quantum efficiency). Consequently, it is not necessary to assume the measurement of a split photon, as in figure 6, and we therefore omit the beam splitter in the following figures 7 and 8.

### 5.2. Lifetimes of the indirect single-photon detector's components

In this section, we calculate how long the components of the indirect single-photon detector shown in figure 7 stay in quantum superposition during detection, where all calculations in this and the following sections are done for room temperature.

For the photodiode in our indirect single-photon detector, we refer to a thick silicon SPAD reported in reference [23] with a breakdown voltage of $V_B = 420\,V$ and an internal resistance of $R_{SPAD} = 500\Omega$. This SPAD detects for an excess bias voltage of $V_E = 15\,V$ photons with wavelengths of $800nm$ with a quantum efficiency of $50\%$. The SPAD is biased with the plate capacitor shown on the left in figure 7 consisting of copper plates and a dielectric of corundum ($Al_2O_3$). The setup contains a resistor with $R = 2k\Omega$ to slow down the discharge of the plate capacitor by the avalanche current of the SPAD. This resistor shall be realized by a rod of doped silicon with $\rho_\Omega = 5\Omega cm$[7] as shown in the figure. To be complete, we also consider the wires connecting these components, which are regarded by the $100cm$ long copper wire at the right.

When an incoming photon is detected, the following time course of the triggered avalanche current is expected [23]:

$$I(t) = \frac{V_E}{R + R_{SPAD}} e^{-\frac{t}{\tau}} \quad , \qquad \text{with } \tau = (R + R_{SPAD})\, C \tag{38}$$

where $C$ is the capacitance of the plate capacitor. For the parameters in figure 7 the time constant $\tau$ is roughly $\tau \approx 1\mu s$. The avalanche current stops when it falls below the latching current of the SPAD of $I_q \approx 0.1mA$ [23], which is at $t_s \approx 3.8\,\mu s$. At this point in time the voltage at the plate capacitor has dropped from $V_B + V_E = 435\,V$ to almost $V_B = 420\,V$, and the square of the current flow until this time is $\int_0^{t_s} dt I(t)^2 \approx 18\mu s(mA)^2$. From the voltage drop at the plate capacitor and this current flow through the SPAD, the resistor and the wire, we get with equation (29) for the thickness change of the plate capacitor's dielectric $\Delta d_{diel}$, and with equations (32) and (33) for the extensions $\Delta r$, $\Delta l$ and $\Delta l_{eff}$ of the SPAD[8], the resistor and the wire the values shown in the middle of figure 7. From these different extensions of the components in the photon-detected and the no-photon-detected state, we get with the results of sections 4.1 and 4.2 for the Diósi-Penrose energies $E_{DP}$ of the plate capacitor and of resistors and wires for the parameter-free and the Diósi-Penrose model with smeared mass-density operator with $T_G = \hbar/E_{DP}$ (equation (1)) the lifetimes $T_G$ of the components[9] shown in the lower part of figure 7. The component with the shortest lifetime, which determines the lifetime of the whole detector, is the plate capacitor, which in case of the parameter-free Diósi-Penrose model can stay for $T_G \approx 93ms$, and in the case of the Diósi-Penrose model with smeared mass-density operator for even $28s$ in superposition.

[7] This corresponds to a n-doping concentration of $\approx 10^{15}/cm^3$.

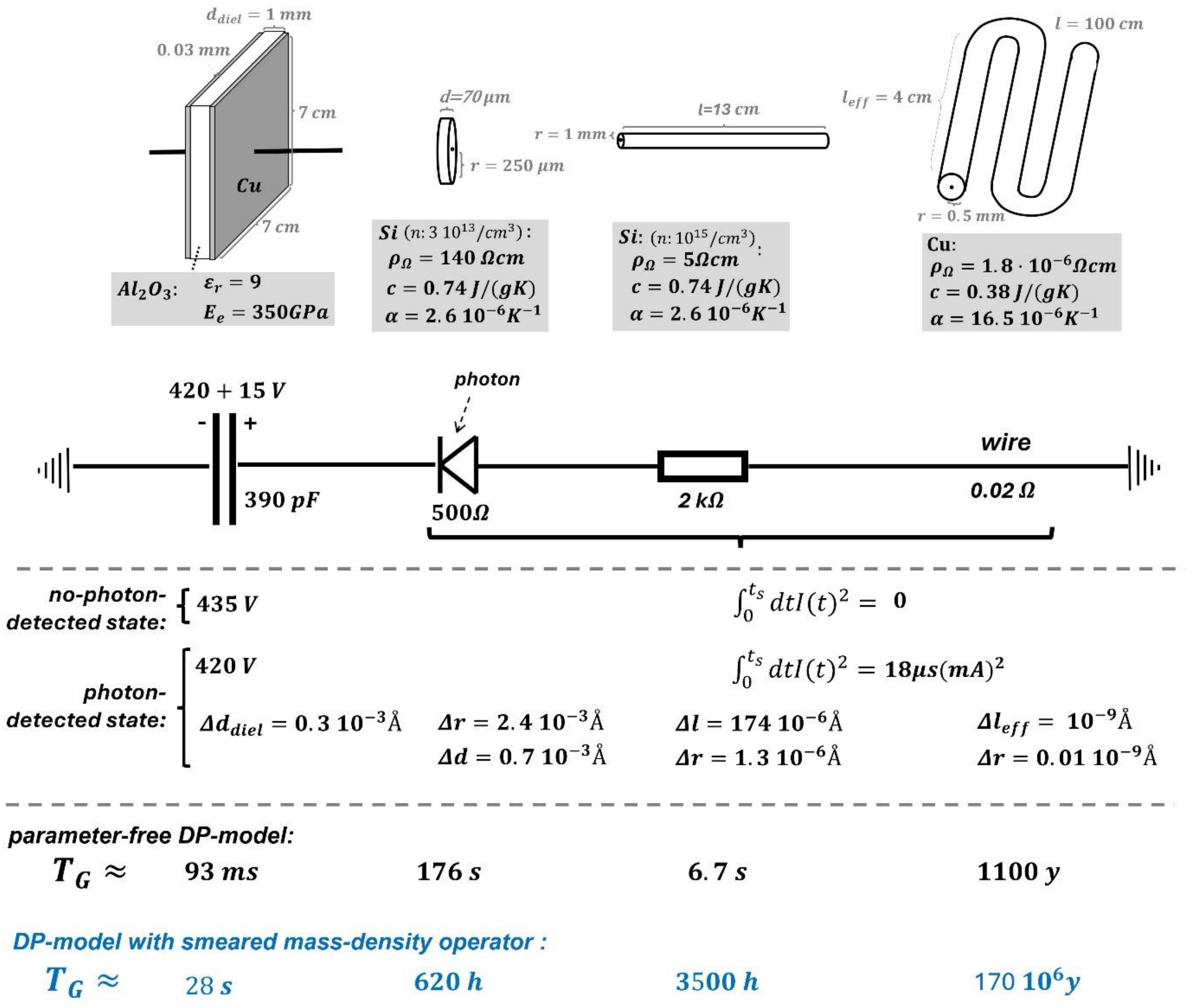

**Figure 7.** *Top:* Components of the indirect single-photon detector.
*Centre:* Differences between the states in which a photon/ no photon is detected.
*Bottom:* Lifetimes of the components for the parameter-free and the Diósi-Penrose model with smeared mass-density operator.

[8] For the thick silicon SPAD we assume an active area with a radius of $r = 250\mu m$ and a thickness of $d = 70\mu m$ [23], which must have a resistivity of $\rho_\Omega \approx 140\ \Omega cm$ (this corresponds to a n-doping concentration of $\approx 3 \cdot 10^{13}/cm^3$) to get a resistance of $R_{SPAD} = 500\Omega$.

[9] For the SPAD in figure 7 (disk with $r = 250\mu m$ and $d = 70\mu m$) the Diósi-Penrose energy is calculated for the parameter-free Diósi-Penrose model with equation (22) for the extended rod, and for the Diósi-Penrose model with smeared mass-density operator with equation (6) for the disk extended to the right.

## 5.3. Transferring a mirror into quantum superposition with an indirect single-photon detector

In this section, we discuss an application of indirect single-photon detectors. Since, as shown in the previous section, indirect detectors can be constructed such that their internal states stay for a long time in superposition, they can be used to deliberately generate superposed states for experiments, provided that the lifetimes of these states are much shorter than that of the detector, so that the influence of the detector on the lifetime of the generated superposition can be neglected. In the following, we show how a small mirror with a mass of $2mg$ can be transferred with help of a piezo-actuator into a superposition in which the mirror is displaced in one state relative to the other. In section 6, we give an outlook on how the generation of such a superposition could be used in a new experiment to measure wavefunction collapse. Here it should be emphasized, as mentioned earlier, that the generated superpositions are decoherent superpositions (improper mixtures), since no effort is made to suppress decoherence between the states.

Figure 8 shows the setup for transferring the $2mg$ mirror into a quantum superposition with a piezo-actuator, which displaces the mirror to the right when it is charged by the avalanche current of the photodiode. For the dimensions shown in the figure, this piezo has a weight of $11mg$. In the following, we calculate exclusively the lifetime of the mirror superposition and assume that the indirect detector is constructed such that its influence on the lifetime of this superposition can be neglected. To simplify the calculation, the capacitance $C$ of the plate capacitor biasing the SPAD shall be much larger than the capacitance of the piezo-actuator $C_{piezo}$. Then the voltage course at the piezo-actuator is:

$$V_{piezo}(t) = V_E(1 - e^{-\frac{t}{\tau_{piezo}}}) \quad . \qquad \text{with: } \tau_{piezo} = R_{SPAD}\, C_{piezo} \tag{39}$$

For the PZT product *PIC 153* of *PI Ceramic GmbH* [24] we get for the dimensions shown in figure 8 and with the relative permittivity $\varepsilon_r = 4200$ of *PIC 153* a capacity of the piezo-actuator of $C_{piezo} = 1300\, pF$ and a time constant of $\tau_{piezo} = 0.66\, \mu s$.

Since the piezo-actuator will shorten the lifetime of the detector so much that it becomes in the order of the settling time of the piezo-actuator of $\tau_{piezo} = 0.66\, \mu s$, the lifetime of the detector cannot be calculated anymore with $T_G = \hbar/E_{DP}$ as in section 5.2, where the lifetimes were much bigger than the detector's settling lime and the Diósi-Penrose energy is then approximately constant over the whole lifetime of the detector. To account for a time-varying Diósi-Penrose energy $E_{DP}$ over the detector's lifetime, we must calculate $T_G$ as follows:

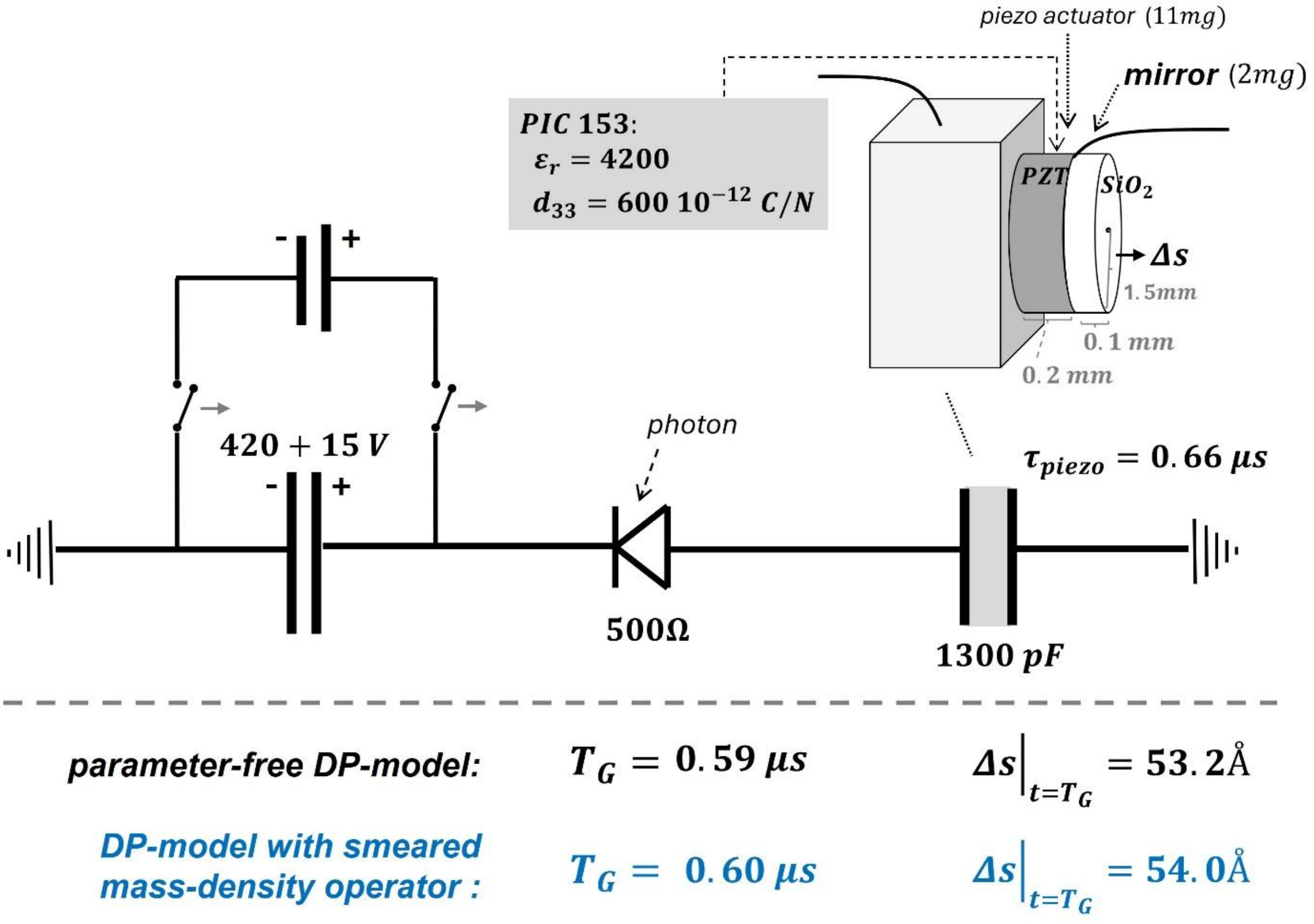

**Figure 8.** *Top:* Setup for transferring a mirror into superposition, where it is displaced in one state by the piezo-actuator, which is charged by the avalanche current of the photodiode.
*Bottom:* Lifetimes of the detector and displacement of the mirror at $t = T_G$ for the parameter-free and the Diósi-Penrose model with smeared mass-density operator.

$$\int_0^{T_G} dt E_{DP}(t) = \hbar \quad . \tag{40}$$

With equation (37) for as disk displaced by a piezo-actuator, which is exactly valid for the Diósi-Penrose model with smeared mass-density operator, we obtain by calculating $\Delta s$ in (37) with equations (36) and (39) and then solving equation (40) a lifetime of $T_G = 0.60\mu s$ and a displacement of the mirror at this time of $\Delta s = 54.0$Å as shown in figure 8. Taking the contribution of the nuclei into account additionally (equation (23) for the parameter-free Diósi-Penrose model, $T_G$ reduces only minimally (see figure 8), since the displacement of 54Å is already larger than ten lattice constants of the used materials (equation (26)).

## 5.4. Control of the indirect single-photon detector's lifetime

The setup proposed in section 5.3 (figure 8) can be used to deliberately control the lifetime of the indirect single-photon detector by replacing the mirror with a solid body of freely selectable size and mass density. In this context, it is often of interest to deliberately shorten the lifetime of the detector, as was done in the experiment in reference [25] testing Bell's inequalities for widely separated partners ($18km$), in which the photon detector also displaced a mass with the help of a piezo-actuator.

In the following, a calculation example is given to show how the lifetime of our detector in figure 8 is shortened when the $SiO_2$ mirror in figure 8, with a thickness of $0.1mm$ and a weight of $2mg$, is replaced by a significantly heavier mass, namely a gold disk with a thickness of $0.2mm$ and a weight of $27\,mg$[10]. As a result, the lifetime of the detector is shortened from $0.6\mu s$ to $T_G \approx 0.2\mu s$. To obtain even shorter lifetimes, SPADs with lower internal resistance can be used. For $R_{SPAD} = 175\Omega$ instead of $500\Omega$ (see figure 7), the lifetime becomes $T_G \approx 0.1\mu s$. Here, the lifetime already reaches the mechanical settling time of the piezo-actuator and the gold disk, which can be estimated from the sound velocities in the materials with $T_{mech} \approx d_{piezo}/v_{\parallel PZT} + d_{disk}/v_{\parallel Au} = 0.11\mu s$[11]. To calculate the lifetime $T_G$ exactly for this case, the formulas for the Diósi–Penrose criterion for superposed solid bodies developed here would need to be further refined.

[10] The mass densities of $SiO_2$ und $Au$ are given in table 1.
[11] For the sound velocities in $PZT$ und $Au$ see table 2.

## 6. Outlook

In this work, two topics have been developed that may be of importance for further research. On the one hand, the detailed collection of formulas for the Diósi–Penrose criterion for superposed solids, and on the other hand, the indirect single-photon detectors.

Since the formulas for the Diósi-Penrose criterion in section 3 were also worked out for very low temperatures, this results are of interest for collapse experiments with solids at low temperatures, as for example the experiment of Marshall *et al.* [26], in which a small mirror is transferred into a superposition of slightly different positions by the light pressure of a single photon.

The analysis of the Diósi-Penrose criterion for the parameter-free and the Diósi-Penrose model with smeared mass density operator has shown that for small displacements on the order of the spatial variation of the solid's nuclei, which is about one tenth of an Ångström at room temperature and roughly a factor of two smaller at low temperatures (see table 1), significant differences arise that may help to identify the correct collapse model in future experiments (see section 3.4)

The analysis of the lifetimes of the detector states in indirect single-photon detectors in section 5.2 led to a surprising result, namely that although the coherence between these states is lost very rapidly, they can nevertheless remain in a superposition for a remarkably long time, up to the order of seconds. Even though this result is initially of purely academic interest, since the superposed states cannot be detected by conventional methods, the possibility of constructing indirect detectors in such a way that they hardly influence the lifetimes of the superpositions they generate (such as the mirror superposition in figure 8) opens up a new approach to measuring wave-function collapse.
In reference [27], it is proposed to place a mirror of a Michelson interferometer into a superposition with a displacement of about 20 Å using an indirect single-photon detector, and to inject and detect single photons in the interferometer. It can be shown that whether the mirror is in a superposition or is in a classical state after reduction affects the probability of detecting photons in the interferometer. What is new about this method is the fact that the photon detection is not disturbed by decoherence between the mirror states, allowing the experiment to be performed even at room temperature. This approach to measuring collapse therefore provides an alternative to the conventional methods based on quantum interference, which are extremely sensitive to decoherence.

## Appendix

In this appendix we derive equation (12) for the spatial variation of the nuclei in a solid using Debye's model for acoustical phonons. Debye's model assumes the following density of states for the acoustical phonons [28]:

$$D_{Deb}(\omega) = 3N \begin{cases} \frac{3\omega^2}{\omega_D^3} & \omega < \omega_D \\ 0 & \omega > \omega_D \end{cases} \quad , \tag{A1}$$

where $N$ is the total number of atoms of the solid and $\omega_D$ the Debye frequency. The Debye frequency is related to the Debye temperature $\Theta_D$ of the solid by $\hbar\omega_D = k_B\,\Theta_D$, where the Debye temperature is determined from the temperature profile of the solid's specific heat. Alternatively, the Debye frequency can be determined from the longitudinal and transverse sound velocities of the solid ($v_\parallel$, $v_\perp$) with [28]:

$$\omega_D^3 = \frac{18\pi^2}{\left(\frac{1}{v_{||}^3} + \frac{2}{v_\perp^3}\right)\bar{g}^3} \quad . \tag{A2}$$

Table 2 compares the Debye frequencies determined from the Debye temperature $\omega_D(\Theta_D)$ with the ones determined from the sound velocities $\omega_D(v_\parallel, v_\perp)$ for the solids in table 1, which differ from each other not more than 20% apart from iron.

The mean energy $< E >$ of a single phonon oscillating with frequency $\omega$ can be estimated with the Bose-Einstein statistic as:

$$< E > = \hbar\omega\left(< n > + \frac{1}{2}\right) \qquad , \textit{with} : < n > = \frac{1}{e^{\frac{\hbar\omega}{k_B T}} - 1} \quad . \tag{A3}$$

The mean square displacement $< x^2 >$ of the atoms oscillating by an excited phonon can be determined from the mean potential energy of the phonon $< E_{pot} > = \frac{c}{2} < x^2 >$ (where $c$ is the spring constant), and the relation $< E > = 2 < E_{pot} >$ between the mean energy and the mean potential energy of the phonon. With the relation $\omega^2 = c/M$, where $M$ is the total oscillating mass, which is given by $M = N\bar{m}$, we get:

$$< x^2 > = \frac{< E >}{N\bar{m}\,\omega^2} \quad . \tag{A4}$$

Taking all excited phonons into account, the mean square displacement of an atom in all three spatial directions is then given by:

| | $\Theta_D[K]$ | $v_\parallel, [\frac{m}{s}]$ | $v_\perp[\frac{m}{s}]$ | $\bar{g}[Å]$ | $\frac{\omega_D(v_\parallel, v_\perp)}{\omega_D(\Theta_D)}$ |
|---|---|---|---|---|---|
| $Al$ | 390 | 6420 | 3040 | 2.55 | 1.23 |
| $Si$ | 692 | 8433 | 5843 | 2.71 | 1.13 |
| $Fe$ | 373 | 5950 | 3240 | 2.28 | 1.49 |
| $Cu$ | 310 | 4760 | 2325 | 2.28 | 1.32 |
| $Pb$ | 87 | 2160 | 700 | 3.12 | 1.08 |
| $Au$ | 178 | 3240 | 1200 | 2.57 | 1.09 |
| $Pt$ | 225 | 3830 | 1680 | 2.47 | 1.23 |
| $SiO_2$ | 523 | 5800 | 3700 | 2.32 | 1.14 |
| $Al_2O_3$ | 980 | 9542 | 4368 | 2.04 | 0.89 |
| $PZT$ | 266 | 4250 | 1900 | 2.42 | 1.20 |

**Table 2.** Comparison of the Debye frequencies calculated from the sound velocities of the solid $\omega_D(v_\parallel, v_\perp)$ with the ones determined from the Debye temperature $\omega_D(\Theta_D)$ (see equations (A2) and (14)).

$$< x^2 + y^2 + z^2 >= \int_0^\infty d\omega D_{Deb}(\omega) \frac{< E >}{N\bar{m}\,\omega^2} \quad . \tag{A5}$$

This mean square displacement in all three spatial directions is related to the spatial variation of the nuclei $\sigma$ by $< x^2 + y^2 + z^2 >= 3\sigma^2$ [12], which leads directly to our result (12).

[12] Note that $\int d^3\boldsymbol{x}\, \boldsymbol{x}^2 \frac{\rho(\boldsymbol{x})}{m} = 3\sigma^2$ with $\rho(\boldsymbol{x})$ according to equation (9).